# Green LTE Broadcasting

Amine Bakhshaie [1], Hamid Shahrokh Shahraki [2]

[1,2] Department of Electrical Engineering, University of Kashan, Kashan, Iran.

[1] bakhshaie.amine@gmail.com

[2] shahrokh@kashanu.ac.ir

*Abstract*— LTE broadcast is a communication standard which is expected to revolutionize the multimedia service provision in the future. The distinctive features of this standard such as increased service capacity, high spectral efficiency, and spectral adjustability have encouraged many companies to select this standard for the future generation broadcasting. In this paper, after a complete introduction of this standard and its unique features, its application in the Heterogeneous Networks (HetNets) is investigated. In this study, application of the LTE broadcast standard under a new scenario called Super cell is examined. Based on the results, the suggested scenario can be considered as one of the best LTE broadcast scenarios. The main aim of this scenario is to decrease the total energy consumption. Therefore, the study has demonstrated that this scenario can reduce the total energy consumption and offers a green servicing by taking advantage of the small cells capability and the D2D communication technology.

## 1. INTRODUCTION

Based on the previous research, the most amount of data traffic in the mobile networks is due to video and multimedia services. Moreover, since the users are interested in receiving these services through smartphones and tablets, it is expected that this traffic should increase over 70% by the year 2020[1].

One of the best proposed standards for service provision to this high volume of data traffic is LTE broadcast. The LTE broadcast standard, which is also called evolved multimedia broadcast multicast service (eMBMS), is indeed an upgraded version of the MBMS standard, which was already developed by 3GPP for the UMTS mobile network[2,3].

Application of the OFDM structure as physical layer in the LTE standard and its progressing versions and the distinctive features of this structure in the SFN networks have caused many companies to use it as a main choice for the future generation of broadcasting. Some of the distinctive features in the design and development of the LTE broadcast are as follows:

- LTE broadcast allows for the change and adjustment of the receiver with respect to the type and volume of the received service. Therefore, it permits the receiver to choose the data reception in unicast or broadcast modes or in a combination of these two modes. This results in the high flexibility of this standard in terms of service capacity and high spectral efficiency.
- LTE broadcast makes it possible for FDD and TDD based systems to be employed in all frequency bands under the LTE coverage.
- All receivers which are equipped with the eMBMS reception technology can receive the LTE broadcast service. This leads to the wide spectral application of the receivers such as smartphones and tablets.
- The geographical range under the LTE broadcast service might change in terms of applications and number of users. This feature provides a wide variety of applications for the LTE broadcast. Some examples are the users' LTE coverage in a stadium, users' LTE service overage in shopping centers, or in a larger geographical limit such as news coverage in a country or provision of software updates in the form of broadcast for all users.

It is worth noting that the LTE broadcast is a completely new and developing idea and many researchers and companies are now working on novel technologies and scenarios for the optimization of this standard.

One of the best defined scenarios for provision of the LTE broadcast services is to employ this standard under the Heterogeneous Networks [4, 5]. In the Heterogeneous networks, a large number of microcells with BTS power are distributed under the coverage of the previous macrocells. The coverage area of these microcells can be the users' coverage in a building (indoor scenario) or the users' coverage in a part of a city with a high traffic of the data demand (outdoor scenario). The use of the HetNets structures for the LTE broadcast servicing brings about many advantages such as:

- Improved performance of the venue casting

When small cells are utilized, the broadcasting process can be centralized more exactly in a defined location. However, if the frequency bands of small cells and macrocells are different, then the broadcasting in the assumed small cell can be performed independently of the macrocell servicing. This leads to a considerable improvement in the capacity and quality of the service provision.

- Improved efficiency in SFN networks

By using small cells and allowing for overlapping their coverage area, the service quality for the users, and especially for those located in the cellular boundaries, considerably increases. Moreover, with regard to the connection of all the BTSs of the small cells through backhaul link, the synchronization problem in the SFN networks can be removed. This leads to a reduction in the number of handover and cut of the connection [6].

- Opportunistic application of the unused frequency bands

Due to the independence of the frequency band of microcells and macrocells, the microcell are allowed to increase the capacity and quality of their services through the opportunistic use of the existing spectral holes.

However, one important aspect which has been considered in the last few years in the emergence and development of new technologies is their amount of effect in the provision of $CO_2$ emission or, better to say, their degree of greenness. With regard to the future horizon of the LTE broadcasting as described above, it is clear that the development of this standard has a substantial share of the $CO_2$ emissions in ICT industry. Therefore, in the selection of a proper scenario for the LTE broadcasting, in addition to the capacity and quality of the network, the consumed energy by the key parts such as user terminals and BTSs, and the main core of the network including different segments connection method and the service management way should also be taken into account[7].

While considering all of the above issues, the present study has also investigated another aspect defined in the LTE-A standard. This feature is D2D communication which allows for the direct connection among the users. In recent years, much attention has been paid to the study of the LTE-based D2D communication from different aspects and perspectives [8-10]. However, its application for the broadcasting can bring about the following merits:

- Traffic offloading

As said before, in the LTE broadcasting view, it is required to transmit a high traffic of data in a limited location. Applying D2D capability to the HetNets can be influential in provision of such a traffic volume.

- Multi hop transmission

On many cases, signal transmission by BTS for the users deployed in the cell boundaries requires a high amount of power consumption, and sometimes it might be impossible to give service to these users due to the badness of the channel conditions. In these cases, application of cooperative D2D communication can not only address the problem of service provision but also reduce the amount of energy consumption.

- Varieties in the uses of the D2D communication in different forms such as in-band or out-band, underlay or overlay, and centralized-based control or distributed-based control have brought about advantages like the proper frequency interference control management and optimal spectral use.

In view of the aforementioned, the main aim of the present study is to propose a proper scenario for the future generation LTE broadcasting with respect to the $CO_2$ emission problem. To this end, first, a scenario called super cell is introduced, which is indeed a three-layer network composed of macro cell, small cell (i.e., microcell), and D2D.

After that, by the investigation of the suggested scenario under the LTE standard, effect of the small cells and cooperative D2D on the total energy consumption of the considered network is evaluated. Our simulation results show that the proposed scenario can be one of the best future generation broadcasting scenarios.

## 2. SUPER CELL ARCHITECTURE

Fig. 1 shows the structure of a cell from the Super cell network for application in the LTE broadcasting. The first part of the structure is similar to the conventional MBMS structures in the macrocell networks in a way that BM-SC block acts as an interface between the content provider and the wireless section and does the process of transformation and preparation for wireless broadcasting.

MBMS-GW block is responsible for management and transmission of the considered data signals among all the BTSs of the macrocell. Note that transmission of similar data to all the SFN cells in the macrocell range (high coverage area) is also a responsibility of this block. Control signals of the macrocells are also supplied by the MME block. These control signals involve all information about synchronization, QOS control required in each cell, and handover management for mobile users in the macrocell scale.

The major part shown in Fig. (1) is related to the interior structure of a given macrocell. The macrocell interior structure has been designed based on the phantom cells theory, in a way that each macrocell can involve a large number of phantom cells [11]. The phantom calls are cells with a small geographical area and separate low power BTSs. Note that in the broadcasting application, any area that involves a great number of users with similar service demands can form a phantom cell.

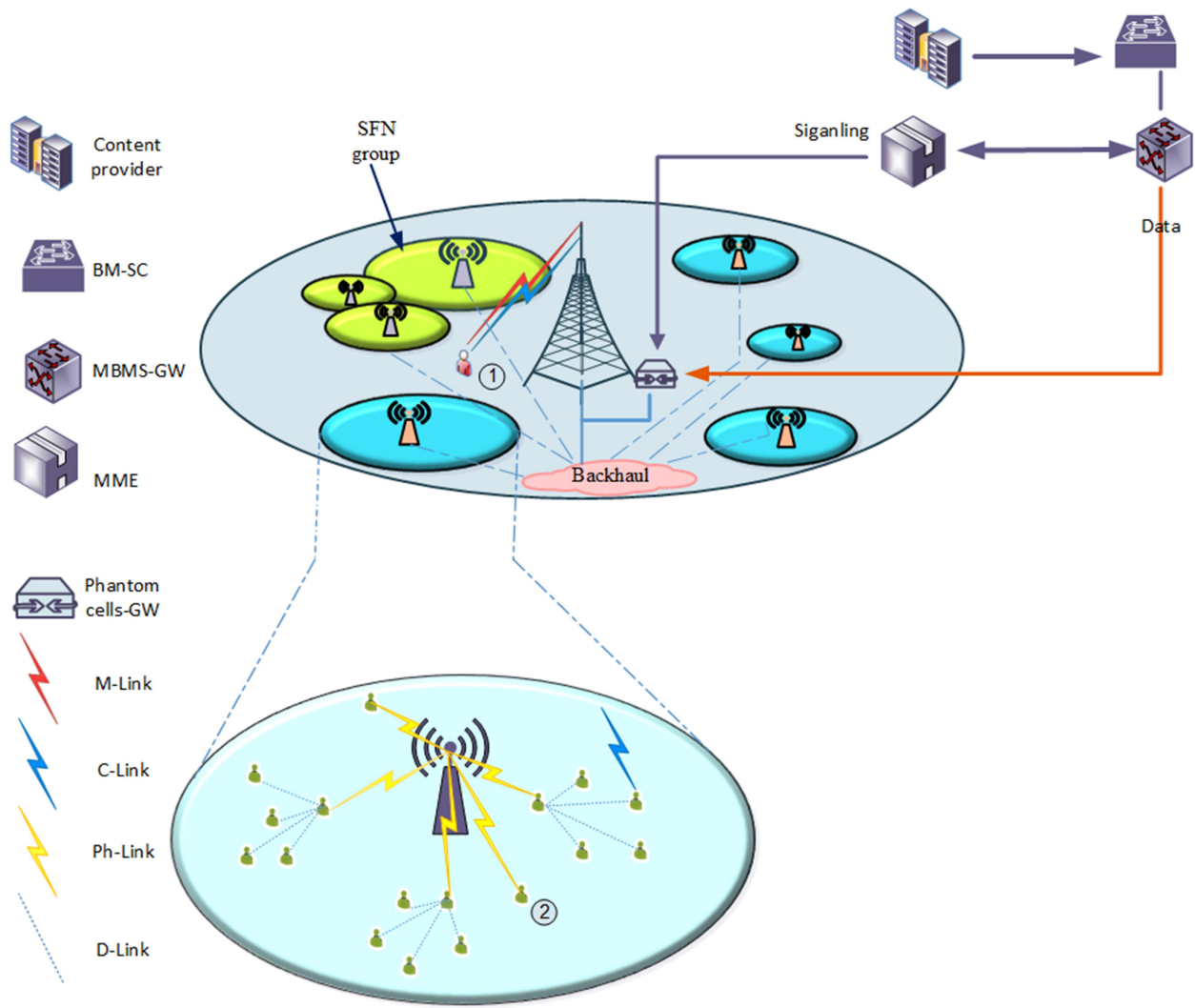

Fig. 1- LTE Broadcast via Super Cell

Phantom cell application can produce many important advantages such as data traffic control and increased system capacity using the backhaul link among all the BTSs of the phantom cells and the BTS of the macrocell. Accordingly, in the structure of each macrocell, a block called phantom cell gateway is considered, which performs the task of transmitting similar data to the all phantom cells with similar demands. Note also that by dividing the phantom cells in each macrocell, a SFN network can be formed in a small geographical scale.

Another particular feature of the phantom cells is that the control signals for all the users in a macrocell's area are directly controlled by macrocell BTS (C-Link). This feature offers such advantage as reduced handover and centralized management of the general network frequency interferences [12].

Note that in the original design of the phantom cells, the users are allowed to receive data by two different frequencies. In other words, the frequency band of a phantom cell is different from the main frequency of the macrocell.

The third layer in the suggested network is related to the internal structure of a phantom cell. As said before, in order to reduce the total energy consumption for giving service to the users of a phantom cell, the D2D communication capability is also utilized. Since for all the users the control link is managed by the BTS of the macrocell, the central station considers the total energy consumption criterion and determines that a user receives the considered data from macrocell BTS, phantom cell BTS, or through the D2D link. Note also that in the D2D data transmission, the working frequency can be equal to the transmission frequency of the corresponding phantom cell or that the transmission is in the form of out-band and conducted in a different frequency.

Details of the algorithm for determination of data transmission method for a given user are given in the next section.

## 3. ENERGY CALCULATION

Assume that the $k^{TH}$ user is to receive a service with a volume of $S_T$. This service can be broadcasted by the macrocell BTS, phantom cell BTS, or by another user.

If the consumed power related to the receiver of this user is shown by $P_{T,k}$, then the total energy consumed by this user for receiving the intended service will be equal to $S_T \cdot \frac{P_{T,k}}{R_k}$, where $R_k$ shows the rate of the data reception by the user.

However, as said before, the given user can receive this service from each of the above mentioned links. If the power consumption of the receiver that broadcasts this service is generally shown by $P_{T,X}$, then the consumed energy by the receiver for transmission of this service will also be equal to $S_T \cdot \frac{P_{T,X}}{R_k}$. Therefore, in order to calculate the total energy and extract the proper algorithm for minimization, the following situations can be taken into account:

**Situation 1:**

The intended service is directly transmitted by the macrocell's BTS. In other words, service provision by small cells and D2D communication is disregarded. In this situation, the total energy consumption is as follows:

$$E_{Total} = \frac{S_T \cdot P_{T,M-L}}{\min R_k} + \sum_{K=1}^{N} \frac{S_T \cdot P_{R,k}}{R_k} \qquad (1)$$

In the above relation, the first term is related to the energy consumed by the macrocell BTS for transmission of the information. In order to ensure that even a user with the worst channel conditions receives the intended service completely, the minimum rate will be considered in the fraction denominator. The second term of this relation shows the sum of the energies consumed by the users' receivers. $N$ shows the total number of users in the range of macrocell, and the information communication link through the macrocell BTS is abbreviated by M-Link.

**Situation 2:**

In this situation, the intended user receives service through the BTS of a phantom cell which is located in its geographical limit. Note that in this situation, data is also broadcasted through the said BTS, and the user receives it directly. In this regard, the total energy consumption involves the energy consumed by the BTSs of the phantom cells, and the total receiver energy of all the users can be expressed as follows:

$$E_{Total} = \sum_{i=1}^{N_{PH}} \frac{S_T . P_{T,PH-Link}}{\min R_{k,i}} + \sum_{i=1}^{N_{PH}} \sum_{K=1}^{N_{i,K}} \frac{S_T . P_{R,k,i}}{R_{k,i}} \qquad (2)$$

In the above relation, $N_{PH}$ shows the total number of the phantom cells existing in the limit of a given macrocell and $N_{i,K}$ shows the number of existing users in the $i^{th}$ phantom cell. Energy calculation is here similar to the explanations in relation (1), except that in this relation $\min R_{k,i}$ represents the rate of a user that has the worst conditions for receiving service in the limit of the $i^{th}$ phantom cell. The information communication link through the phantom cell BTS is abbreviated by PH-Link.

**Situation 3:**

In this situation, the users in the space of a phantom cell can receive service cooperatively and via the D2D link. To this end, the total users are clustered with a defined criterion. For each class or cluster, a user is selected as the head of the cluster.

Usually, the user with the least distance to the BTS, or equivalently the user with the best channel conditions for data transmission, is presumed to be the head of the cluster. A user which plays the role of the cluster head in a cluster receives the intended data from the BTS of the phantom cells and then multicasts these data to all of the users in its cluster. In this situation, the total energy consumption involves three sections and can be calculated as follows:

$$E_{Total} = \sum_{i=1}^{N_{PH}} \frac{S_T \cdot P_{T,PH-Li}}{\min R_{k,i}} + \sum_{i=1}^{N_{PH}} \sum_{j=1}^{N_{i,C}} \frac{S_T \cdot P_{T,D-L}}{\min R_{k,ij}} + \sum_{i=1}^{N_{PH}} \sum_{j=1}^{N_C} \sum_{K=1}^{N_{J,K}} \frac{S_T \cdot P_{R,k,ij}}{R_{k,ij}} \qquad (3)$$

In the above relation, the first term shows the amount of energy consumed by the BTS of the phantom cells for transmitting information to the cluster heads of the considered clusters. The second term is related to the energy consumed by a cluster head for broadcasting information among the users of its own cluster. $N_{i,C}$ shows the number of clusters in the limit of the $i^{th}$ phantom cell. Note that in the denominator of the second term, in order to ensure information transmission, the minimum value is devoted to the user with the worst conditions in relation to the cluster head of the corresponding cluster.

Finally, the third term represents the total energy consumed by the receivers of the users. Information communication link from a cluster head to the users in its cluster is abbreviated by D-Link.

## 4. PROPOSED ALGORITHM

The suggested algorithm is based on the assumption that all the users of a phantom cell are going to receive a service with the same content.

**Designating BTS**

- First, the total energy consumption related to the service reception for all the users by the macrocell BTS is calculated.
- For each user, if the energy of receiving service from each BTS of the phantom cell is lower than the energy of receiving service from the macrocell BTS, then that user will be designated to the BTS of that phantom cell. In this way, the users in the limit of each phantom cell are determined. Other users receive their desired service directly from the macrocell BTS (like user No. 1 in Fig. 1).

**Clustering the users of each phantom cell:**

- For all the users of a phantom cell, energy of receiving service is directly calculated from the corresponding BTS.
- The user with the best condition in terms of the rate of reception is selected as the cluster head.
- For each user in the limit of phasntom cell, if the energy of service received from the cluster head user is lower than the direct reception energy, then that user will be placed in the cluster of that cluster head user.
- For users which are not placed in that cluster, the processes of cluster head selection and new clustering are repeated.
- This will continue until all the users are clustered in the phantom cells.

Note that among the users in a given phantom cell limit, some users might exist that are not subsumed under any clusters and receive information directly from the BTS of the phantom cell (like user No. 2 in Fig. 1). Moreover, since in the defined scenario, the control link of all the users is controlled by the macrocell BTS, the management of the service reception method for each user is performed by this link.

## 5. SIMULATION RESULTS

In this section, by simulation of the suggested scenario and study of the obtained results, the optimization of our scenario in terms of energy consumption is investigated. This simulation has been performed for a macrocell with 10 phantom cells. Each phantom cell indeed involves a series of users with the same service demand that are placed in the geographical range of the BTSof the phantom cell.

Location of the users of each phantom cell has been randomly considered. For performance evaluation in terms of energy consumption, the service provision to the users has been simulated under three different situations. In the first situation, without considering the HetNet structure, all the users are directly serviced by the macrocell BTS.

In the second situation, service is provided by the BTSs of the phantom cells. Note that in this situation, information about the users of each phantom cell is transmitted to the phantom cell BTS via the backhaul link, and the related BTS will broadcast this service to all the users.

In the third situation, D2D communication capability is added to the scenario and the users in the limit of any given phantom cell are serviced with proper clustering according to the proposed algorithm.

For all these three communication links, a random channel with a Rayleigh distribution is considered, while the model of path drop for each link is different and given in Table (1). Other simulation parameters are also summarized in Table (1).

Table 1. Simulation parameters

| | Parameter | Value | | Parameter | Value |
|---|---|---|---|---|---|
| Simulation parameters | Macro cell radius | 500 m | Simulation parameters | $P_{R,M-Link}$ | 1.8 Joul/s |
| | Phantom cell radius | 50 m | | $P_{R,Ph-Link}$ | 1.2 Joul/s |
| | Number of phantom cell | 10 | | $P_{R,D-Link}$ | 0.9 Joul/s |
| | Total bandwidth | 10 MHz | | Noise power | -147 dbm/Hz |
| | $S_T$ | 1 G bits | | Shadowing variance | 8db |
| | $P_{T,M-Link}$ | 40 Joul/s | Pass Loss Model | Macro cell | $128 + 37.6 \, log_{10}(d)$ |
| | $P_{T,Ph-Link}$ | 10 Joul/s | | Phantom cell | $37 + 20 log_{10}(d)$ |
| | $P_{T,D-Link}$ | 0.125 Joul/s | | D2D | $42 + 16.9 \, log_{10}(d)$ |

As shown in the diagrams of Fig. (2), the use of the phantom cell architecture for multimedia servicing offers a considerable reduction in the energy consumption. It should be mentioned that a major part of this reduction in the energy consumption is related to the consumed energy by the users' receivers. In this context, not only is the total considered structure green, but also the useful life of the receivers' batteries increases.

Fig. (2) also shows the effect of giving service to the receivers of a Super cell through clustering. By investigation of the diagrams, it is observed that a combinational use of phantom cell and D2D communication techniques offers the best efficiency in terms of energy consumption.

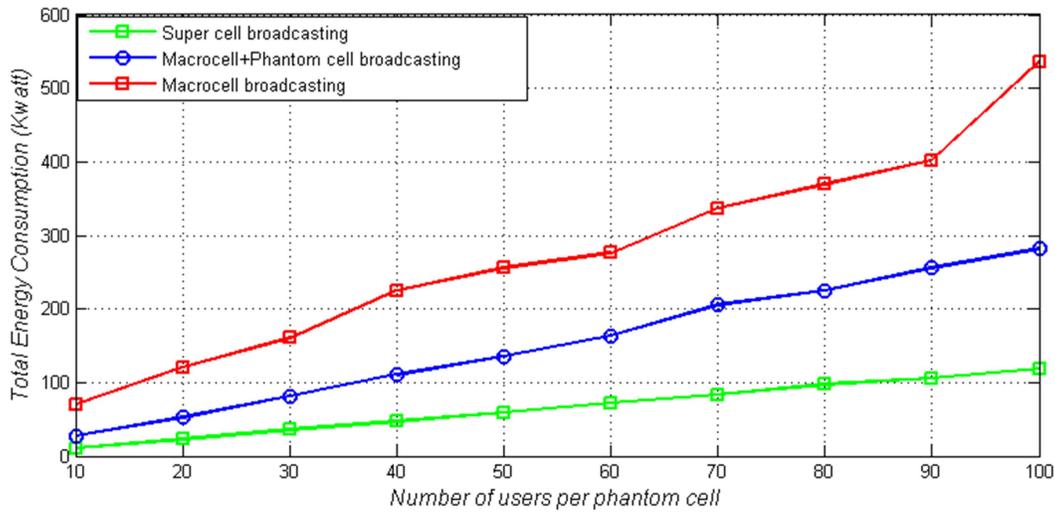

Fig. 2- Comparison of the Total power consumption based on the number of users under different scenarios

Clearly, the value of the energy efficiency depends on the number of the phantom cells, users distribution in the limit of each phantom cell, and other parameters such as the applied frequency band for each of the aforementioned transmissions. With respect to the versatility of our proposed scenario concerning the above issues, results of Fig. (2) can thus be assumed as an example confirming that our suggested scenario is optimized for reduction of energy consumption.

## 6. CONCLUSION

In this work, Super cell scenario has been introduced for the future generation broadcasting application. By investigation of the different aspects of the LTE broadcasting, the use of this standard under the proposed scenario has been considered. Besides, by offering a proper algorithm for reduction of energy consumption, it has been shown that this scenario can be one of the best choices for multimedia servicing of the future generations.